\documentclass[twocolumn,pra,amsmath,amssymb,superscriptaddress,longbibliography]{revtex4-1}
\usepackage[utf8]{inputenc}
\usepackage{amsfonts,amsmath,amssymb}
\usepackage{graphicx}
\usepackage{bm} % boldmath
\usepackage{mathrsfs}
\usepackage{subfigure}
\usepackage{textcomp}
\usepackage{hyperref}
\usepackage[flushleft]{threeparttable}
\usepackage[per-mode=symbol]{siunitx}
\usepackage[version=3]{mhchem}
\def\qr{{\bf r}}                                   % fettes r
\def\qk{{\bf k}}                                   % fettes k
\begin{document}

%\title{Phases of a microwave-shielded Bose-Einstein condensate of molecules}
\title{Phases of a Bose-Einstein condensate of microwave-shielded dipolar molecules}

\author{Chiara J. Polterauer}
\author{Robert E. Zillich}
\affiliation{Institute for Theoretical Physics, Johannes Kepler University Linz, Altenberger Straße 69, 4040 Linz, Austria}

\date{\today}

\begin{abstract}
Bose-Einstein condensation of dipolar molecules can be achieved by shielding loss channels with microwave fields.
The microwave coupling can be approximated by effective dipole-dipole interactions with a short-range repulsion.
We study properties and stability of these molecular Bose gases with a many-body variational method,
the hypernetted-chain Euler-Lagrange method for a wide range of densities and repulsion strengths
of the microwave shield. We find a homogeneous gas-like phase which, however, is unstable at low density
against density waves: at a critical density, which depends on the repulsion strength,
the dipolar fluid undergoes a phase transition to a layer phase. Thus, if the molecular condensate is
expanded adiabatically by decreasing the confinement strength,
it will spontaneously form layers at the critical density.
These quasi-two-dimensional layers can be self-bound, hence form two-dimensional liquids.
By varying the microwave shield, the predicted equilibrium densities span more than an order of magnitude.
\end{abstract}

\maketitle

\section{Introduction}

In ultracold molecular quantum gases, studied for
many years\cite{koehlerRMP06,niScience08,deiglmayrPRL08,takekoshiPRA12}, the molecule rotation provides a
handle to manipulate the effective interaction\cite{buechlerPRL07,baranovChemPhys12}, and thus the many-body
state, and realize effective spin Hamiltonians\cite{micheliNaturePhys06,gorshkovPRA11}.
Considerable progress in cooling and controlling molecular quantum gases has been achieved\cite{langenNatPhys24},
in particular in optical lattices\cite{ospelkausPRL06}, where
molecules can be isolated from each other and only internal degrees of freedom are
coupled via dipole-dipole interactions. Without an optical lattice, the abundance of scattering channels
generally leads to high two- and three-body losses. One way to overcome this problem is microwave shielding
\cite{karmanPRL18,karmanPRA19,lassablierePRL18,dengPRL23,karmanPRX25}:
dressing rotational states with microwaves generates effective
dipole-dipole interactions, and a Bose-Einstein condensate (BEC) of NaCs molecules
has indeed been realized with this method\cite{bigagliNature24}.

The possibility of studying effects of the long-ranged anisotropic dipole-dipole interaction
in a quantum system is not new\cite{chomaz_dipolar_2023}. BECs
of atoms with magnetic dipole moment\cite{goralPRA00} have been realized with
Cr\cite{griesmaierPRL05,lahayeNature07}, Er\cite{aikawaPRL12,chomazPRX16},
and Dy\cite{luPRL11dysprosium,Kadau2016}, and a wealth of phenomena caused
by dipole interactions have been proposed and found, such as roton excitations
\cite{Santos2003,odellPRL03,chomazNatPhys18,Natale2019,Schmidt2021,Blakie2020},
self-bound droplets\cite{chomazPRX16,maciaPRL16,Schmitt2016,Ferrier2016}, or supersolid behavior
\cite{Leonard2017,Tanzi2019supersolid,Zhang2019,Roccuzzo2019,Hertkorn2021,Tanzi2021science,norciaNature21}. 

From the theoretical side, many phenomena in quantum gases with magnetic dipole moments can
be described, at least approximately, by perturbations to the mean field approximation,
such as Lee-Huang-Yang corrections\cite{limaPRA11,limaPRA12}.
But due to the much larger electric dipole moment of heteronuclear molecules compared
to the magnetic dipole moment of atoms, mean field based approximations are no longer applicable
to molecular quantum gases\cite{langenPRL25,jinPRL25}.
Correlations can no longer be treated as perturbation of the mean field result (i.e.\ as quantum fluctuations
of the classical field describing the Bose condensate). We have to use
non-perturbative many-body methods to account for correlations. Exact results for bosons
can be obtained with quantum Monte Carlo simulations
\cite{astraPRL07,dipolePRL09,maciaPRA11,maciaPRL12,maciaPRL16,bombinPRL17,Bottcher2019droplet,Kora2019,langenPRL25},
which come with a considerable computational cost. More efficient, but approximate, are
many-body methods such as variational methods.

Like a gas of atoms with magnetic moments, a gas of dipolar molecules can be expected to form self-bound states,
i.e. liquid droplets, if the attractive part of the dipole-dipole interaction is sufficiently strong. Indeed,
Monte Carlo simulations have recently predicted pancake-shaped self-bound droplets of microwave-shielded
dipolar molecules\cite{langenPRL25}, above a critical number of molecules $N$, depending on the
strength of the microwave shield.
In this work we systematically study the phases -- self-bound liquid or gas-like phase -- of
microwave-shielded dipolar molecules in the thermodynamic limit $N\to\infty$.
We use the hypernetted-chain Euler-Lagrange (HNC-EL) method \cite{Kro86,KroTrieste,QMBT00Polls}, a non-perturbative variational many-body
method. HNC-EL has long been used as an efficient alternative to exact MC simulations and is especially
useful if a large parameter space (density, interaction parameters, etc) has to be searched.

The paper is organized as follows: in section \ref{sec:theory} we first review the HNC-EL method for a
homogeneous but anisotropic quantum fluid (section \ref{ssec:theory3D}), and then we derive an approximate 
treatment of thin layers of a molecular BEC, where translation invariance is broken in one direction
(section \ref{ssec:theory2D}). In section \ref{sec:results}, we present our results for the homogeneous
gas-like phase (section \ref{ssec:result3D}) and for self-bound layers (section \ref{ssec:result2D}).
We summarize our findings in section \ref{sec:conclusion}, and propose future directions of investigation
in section \ref{sec:outlook}.

\section{Theory}
\label{sec:theory}

We give a brief introduction to the HNC-EL method for a three-dimensional dipolar BEC and to
our modifications for self-bound layers. Both cases are modeled by the Hamiltonian for $N$ interacting particles
\begin{align}
  H = -{\hbar^2\over 2m}\sum_{i=1}^N\nabla^2_i + \sum_{i<j} v(\qr_i-\qr_j)
  \label{eq:H}
\end{align}
where $v(\qr_i-\qr_j)$ is the interaction potential between particle $i$ and $j$,
in our case the effective dipolar interaction generated by the a single microwave field
\cite{dengPRL23}. Note that we do not include 
an external potential because we are particularly interested in self-trapped phases.
The Hamiltonian (\ref{eq:H}) is translationally invariant, therefore we study 
homogeneous phases in the thermodynamic limit $N\to\infty$ and $V\to\infty$ with
finite $N/V=\rho$.

\subsection{Three-dimensional homogeneous anisotropic fluid}
\label{ssec:theory3D}

\begin{figure}[hbtp]
    \centering
    \includegraphics[width=1.0\linewidth]{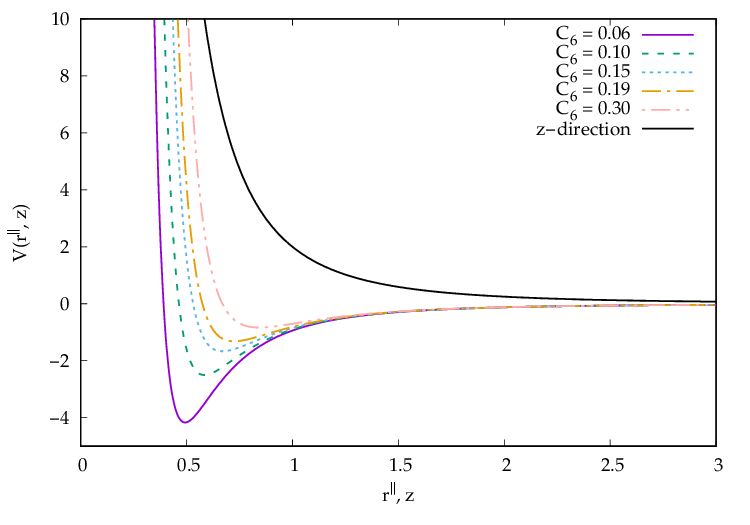}
    \caption{The microwave-shielded dipolar interaction $v(\qr)$ for (see eq.~(\ref{eq:vbare})
    in the $xy$-plane as a function of
    $r^\|=\sqrt{x^2+y^2}$ and in $z$-direction, for different repulsion strengths $C_6$.}
    \label{fig:pot3D}
\end{figure}

Coupling the lowest rotational states of dipolar molecules with a microwave field creates an effective
dipole interaction with a short-ranged repulsion\cite{dengPRL23} that shields the molecules
against losses due to the strong attraction of the pure dipole interaction,
e.g.\ of two dipoles oriented head to tail.
For elliptic polarization, the dipole interaction is anisotropic in all three directions of space, but
in this work, we restrict ourselves to the cylinder-symmetric case
%\begin{align}
%  v(\qr) =\frac{C_6}{r^6} \sin^2{\theta} \{1 - F^2_{\xi}(\varphi) +  [1 - F_{\xi}(\varphi)]^2\cos^2(\theta)\} \nonumber \\ 
%  + \frac{C_3}{r^3}[3\cos^2(\theta) - 1 + 3F_{\xi}(\varphi)\sin(\theta)]
%  \label{eq:vbare}
%\end{align}
$v'(\qr) = \frac{C'_6}{r^6} \sin^2{\theta} (1+\cos^2(\theta)) + \frac{C_3}{r^3}(3\cos^2(\theta) - 1)$
where $C_3$ and $C'_6$ can be adjusted by the Rabi frequency and the detuning\cite{dengPRL23}.

We use the dipole length $r_0={C_3 m\over \hbar^2}$ as unit of length and
and $E_0={\hbar\over m r_0^2}$ as unit of energy. This eliminates $C_3$ from the
interaction, which becomes
\begin{align}
    v(\qr) = \frac{C_6}{(r/r_0)^6} \sin^2{\theta} (1+\cos^2(\theta)) + \frac{3\cos^2(\theta) - 1}{(r/r_0)^3}
    \label{eq:vbare}
\end{align}
with $C_6={C'_6\over C_3^4}\big({\hbar^2\over m}\big)^3$.
Fig.\ref{fig:pot3D} shows $v(r^\|,z)$ in the $xy$-plane as function of $r^\|=\sqrt{x^2+y^2}$
and in $z$-direction for the range of $C_6$ values relevant for our work. Note that the microwave-shielded
molecules attract each other in the $xy$-plane and repel each other in $z$-direction -- the opposite of the
behavior of unshielded polarized dipoles.

The Jastrow ansatz \cite{Jastrow55} for the ground state for a homogeneous Bose fluid is
\begin{align}
  \Psi(\qr_1,\dots,\qr_N) = \exp\Big[ {1\over2} \sum_{i<j} u(\qr_i-\qr_j)\Big]
  \label{eq:Psi}
\end{align}
where $u(\qr)$ are (anisotropic) pair-correlations.
For strong interactions and high densities, three-body correlations $u_3(\qr_1,\qr_2,\qr_3)$ should
be included to achieve good agreement with exact Monte Carlo results~\cite{Kro86}. However, the
treatment of $u_3(\qr_1,\qr_2,\qr_3)$ would be numerically much more expensive and defeats the purpose
of scanning the parameter space $(\rho,C_6,)$ over several orders of magnitudes efficiently.
Extensive experience with the HNC-EL method applied to superfluid helium-4 has shown that
even if quantitative agreement for energy and structural quantities is not
achieved with pair correlations $u(\qr)$ only, the qualitative behavior and trends
are well captured by $u(\qr)$ even for a strongly correlated quantum fluid like $^4$He.

Given the variational ansatz (\ref{eq:Psi}), we can calculate the energy per particle
\begin{align}
  e &= {1\over N}{\langle\Psi|H|\Psi\rangle\over\langle\Psi|\Psi\rangle} = \bar t + \bar v
\label{eq:E}
\end{align}
with the kinetic energy per particle
\begin{align}
  \bar t & = \frac{\rho}{2} \frac{\hbar}{m} \int\! d^3r\, |\nabla \sqrt{g(\qr)}|^2 \nonumber\\
    & - \frac{\hbar^2}{4m} \int\! \frac{d^3k}{(2\pi)^2\rho}\, k^2 (S(\qk)-1)\Big(S(\qk)-2+\frac{1}{S(\qk)}\Big)
\label{eq:T}
\end{align}
and the potential energy per particle
\begin{align}
  \bar v =  \frac{\rho}{2} \int\! d^3r\, g(\qr) v(\qr)
  \label{eq:V}
\end{align}
$g(\qr)$ is the pair distribution function (also called radial distribution function
in the isotropic case)
\begin{align}
  g(\qr) = {1\over\rho^2}{1\over\langle\Psi|\Psi\rangle}
  \Big\langle\Psi\Big| \sum_{i\ne j}\delta(\qr_i-\qr_j-\qr) \Big|\Psi\Big\rangle
\label{eq:defg}
\end{align}
$S(\qk)$ is the static structure factor, which is another measure for the pair correlations since
it is basically the Fourier transform of $g(\qr)$,
\begin{align}
  S(\qk)  = 1 + \rho\int d^3r\, e^{-i\qk\qr}[g(\qr)-1]
\label{eq:defS}
\end{align}
Particularly, a peak in $S(\qk)$ is an indication of an oscillation in $g(\qr)$.
The longer-ranged the oscillation, the higher the peak becomes, until it becomes a Bragg
peak in a solid.
The long-wave length limit of $S(\qk)$ is related to the speed of sound $c$ \cite{Feenberg}
\begin{align}
    S(k) \to {\hbar k\over 2mc}\quad\mbox{for}\ \ k\to 0
    \label{eq:Skzero}
\end{align}

According to Ritz' variational principle, the optimal correlations $u(\qr)$ are obtained by solving
the Euler-Lagrange (EL) equation
\begin{align}
  {\delta e\over\delta u(\qr)} &= 0
  \label{eq:dEdu}
\end{align}
For the calculation of $e$, and hence for solving the EL equation (\ref{eq:dEdu}) we need
a relation between $g(\qr)$ and $u(\qr)$. There is no exact relation, but
a systematic family of approximations has been developed by exploiting a formal equivalence
between the normalization integral $\langle\Psi|\Psi\rangle$ and the classical canonical partition
function at temperature $T$ for a pair-wise interaction $-k T u(\qr)$.
Integral equation theories have been developed to calculate expectation values
by expansion in Mayer cluster diagrams\cite{Hansen}, using the well-established
hypernetted-chain (HNC) relation which is the closure that expresses $g(\qr)$ in terms of $u(\qr)$
\begin{align}
    g(\qr) = \exp\Big[u(\qr) + N(\qr) + B(\qr)\Big]
    \label{eq:HNC}
\end{align}
$N(\qr)$ is the indirect correlation function which can be readily calculated from $g(\qr)$
by solving the Ornstein-Zernicke relation\cite{HaM76,Kro86}. $B(\qr)$ is the infinite sum of the
so-called elementary (or ``bridge'') diagrams, which need to be approximated by truncation
of this infinite sum. For $^4$He, the influence of elementary diagrams on the results have
been investigated in much detail\cite{Kro86}. Usually, only the few lowest diagrams of $B(\qr)$
can be evaluated.

When we use the HNC relation (\ref{eq:HNC}) in the EL equation (\ref{eq:dEdu})
the method is called hypernetted-chain Euler-Lagrange (HNC-EL) method.
Detailed reviews of the HNC-EL method and how to solve the HNC-EL equation (\ref{eq:dEdu})
efficiently can be found in \cite{Kro86,KroTrieste,QMBT00Polls}.

The simplest approximation, HNC0, completely neglects $B(\qr)$.
This approximation is good for low density $\rho$, but we cannot expect quantitative
agreement with exact results for higher $\rho$. Anyway, at higher densities, at least three-body
correlations $u_3(\qr_1,\qr_2,\qr_3)$ need to be included in the variational ansatz for quantitative
agreement, as mentioned above. For the same
reasons why we neglect $u_3$, we also neglect $B(\qr)$ in this work, i.e., we employ the HNC0
approximation. The EL equation, formulated for the observable $g(\qr)$ rather than for $u(\qr)$,
becomes quite simple:
\begin{align}
    -{\hbar^2\over m}\nabla^2 \sqrt{g(\qr)} + v(\qr)\sqrt{g(\qr)} + w_I(\qr)\sqrt{g(\qr)} = 0
    \label{eq:EL}
\end{align}
where $w_I(\qr)$ is the Fourier transform of
\begin{align}
  w_I(\qk) = - \frac{\hbar^2 k^2}{4m} \biggl ( 1 - \frac{1}{S(\mathbf{k})} \biggr )^2 [2S(\mathbf{k}) + 1]
\end{align}

The EL equation (\ref{eq:EL}) has the form of a Schr\"odinger equation for a two-body scattering
problem at zero-energy, where the square root of the pair distribution function $g(\qr)$ takes the role
of the relative two-body wave function and the interaction consists of the actual
interaction $v(\qr)$ and another interaction $w_I(\qr)$ which contains the many-body
physics. $w_I(\qr)$ describes the induced interaction between two particles mediated
by the many-body excitations. One can show that $w_I\to 0$ for vanishing density $\rho\to 0$,
as expected.

\subsection{Quasi-two-dimensional homogeneous fluid}
\label{ssec:theory2D}

As our results will show, a 3D molecular gas can contract into quasi-2D layers,
which breaks translational
invariance perpendicular to the layer plane, taken to be the $z$-direction. The pair
correlation become functions of three coordinates,
$u(|\qr_1^\|-\qr_2^\||,z_1,z_2)$, where $\qr^\|=(x,y)$ is the projection of $\qr$
on the $xy$-plane. The HNC-EL method has been generalized for this latter case and applied
to adsorbed $^4$He films~\cite{EpKro,Clements93,clementsPRB96} and
dipolar layers~\cite{hufnaglJLTP10,hufnaglPRL11,hufnaglPRA13}.

Here we take a much simpler approach, which is valid as long as the thickness of a layer is
less than the average interparticle spacing.  For such a thin layer, we can approximate
the pair correlations by only considering their dependence on the in-layer distance,
$u(\qr_1^\|-\qr_2^\|)$, while using a simple mean-field ansatz for the $z$-dependence
\begin{align}
  \Psi(\qr_1,\dots,\qr_N) = \prod_i\phi(z_i)\ \exp\Big[ {1\over2} \sum_{i<j} u(\qr_i^\|-\qr_j^\|)\Big]
  \label{eq:Psi2D}
\end{align}
As a further simplification, we approximate the one-body functions by Gaussians
\begin{align}
    \phi(z) = {1\over (\pi \sigma^2)^{1/4}}e^{-z^2/2\sigma^2}
    \label{eq:phi}
\end{align}
where the width of the layer is given by $\sigma$.
Instead of a full functional minimization with respect to $\phi(z)$ we only need to
minimize with respect to $\sigma$. Hence we need to solve the equations
\begin{align}
  {de\over d\sigma} &= 0
  \label{eq:dEdsigma}\\
  {\delta e\over\delta u(\qr^\|)} &= 0
  \label{eq:dEdu2D}
\end{align}
Although the parallel coordinates $\qr_i^\|$ and perpendicular coordinates $z_i$
are decoupled in the ansatz (\ref{eq:Psi2D}), these two Euler-Lagrange equation are coupled.
The second equation (\ref{eq:dEdu2D}) can again be brought in the form (\ref{eq:EL}), i.e. the 2D version of
the HNC0-EL equation, but with the bare interaction (\ref{eq:vbare}) being replaced by a
Gaussian-averaged interaction,
\begin{align}
  v^{\rm 2D}(\qr^\|)
    &= \int dz_1 dz_2 \, |\phi(z_1)|^2 |\phi_1(z_2)|^2 v(\qr_1-\qr_2) \\
    &= \int dz \, {1\over \sqrt{2\pi}\sigma} e^{-z^2/2\sigma^2}  v(\qr^\|,z)
  \label{eq:vgauss}
\end{align}
where $v(\qr^\|,z)=v(\qr)$.
In addition the self-trapping in $z$-direction described by (\ref{eq:phi}) leads to a
kinetic energy penalty per particle, given by ${{\hbar^2\over 4m\sigma^2}}$.

Eqns.~(\ref{eq:dEdsigma}) and (\ref{eq:dEdu2D}) have to be solved self-consistently,
where eq.~(\ref{eq:dEdsigma}) can be written as
\begin{align}
    \sigma = \sqrt{\frac{\frac{\hbar^2}{2m}}{\bar v_2-\bar v}}
    \label{eq:dEdsigma2}
\end{align}
$\bar v$ is the expectation value of the interaction potential per particle (\ref{eq:V}),
but in two dimensions and with $v$ replaced by $v^{\rm 2D}$.
$\bar v_2$ is the variance of the interaction in $z$-direction,
\begin{align}
  \bar v_2 =  \frac{\rho^{\rm 2D}}{2}\! \int\! d^2r^\|\, g(\qr^\|)
              \int dz\, {1\over \sqrt{2\pi}\sigma} e^{-z^2/2\sigma^2} {z^2\over \sigma^2} v(\qr^\|,z)
  \label{eq:V2}
\end{align}
Note that the right hand side of eq.(\ref{eq:dEdsigma2}) also depends on $\sigma$.

\section{Results}
\label{sec:results}

We study the microwave-shielded dipolar BEC both in a three dimensional homogeneous phase
(section~\ref{ssec:result3D}) and in a layer phase (section~\ref{ssec:result2D}),
where translational invariance is broken in one spatial dimension. We calculate ground state energies for a wide
range of repulsion strengths $C_6$ and densities $\rho$, as well as structural quantities, the pair
distribution function $g(\qr)$ and the static structure factor $S(\qk)$. Both are central to the
HNC-EL method, but are also important physical observables to characterize a many-body ground state.
We will see that $S(\qk)$ is crucial to understand the stability of the molecular BEC.

Although hard to measure experimentally, the ground state energy per particle $e={E\over N}$ is a useful
quantity to get the big picture of the state of a quantum fluid -- whether it's a liquid or a gas.
From the dependence of $e$ on the density $\rho$, i.e.\ the
zero-temperature equation of state $e(\rho)$, we can distinguish a gas from a liquid.
A gas minimizes its energy by expanding in the absence of a confining potential towards $\rho\to 0$,
therefore $e(\rho)$ falls monotonically towards zero with $\rho$.
A liquid minimizes its energy adjusting its density
until reaching equilibrium density $\rho_{\rm eq}$, even without confinement, i.e.\ at zero pressure.
In other words, a liquid is self-bound.  Thus we can distinguish liquid from gas phase by
simply looking for a local minimum of $e(\rho)$. Additionally, in all know instances,
$e(\rho)$ is negative near $\rho_{\rm eq}$.

\subsection{Three-dimensional homogeneous phase}
\label{ssec:result3D}

We present our results for the energy and structure of a molecular BEC with microwave shielded
dipolar interactions (\ref{eq:vbare}) in three dimensions. Like $v(r^\|,z)$,
the pair distribution function depends only on the two coordinates, $g(\qr)=g(r^\|,z)$,
where $\qr=\qr_1-\qr_2$; the same holds for the static structure factor $S(\qk)=S(k^\|,k_z)$.
$\qr^\|$ and $\qk^\|$ are the projection of $\qr$ and $\qk$ on the $xy$-plane, respectively.
We vary the number density $\rho^{\rm 3D}$ and the repulsion strength $C_6$, with an
emphasize on values of $C_6$ equal and larger than the largest $C_6$ studied in
Ref.~\cite{langenPRL25}, i.e.\ for weaker attraction between the molecules,
because we want to explore possible gas phase regimes.

\begin{figure}[hbtp]
    \centering
    \includegraphics[width=1.0\linewidth]{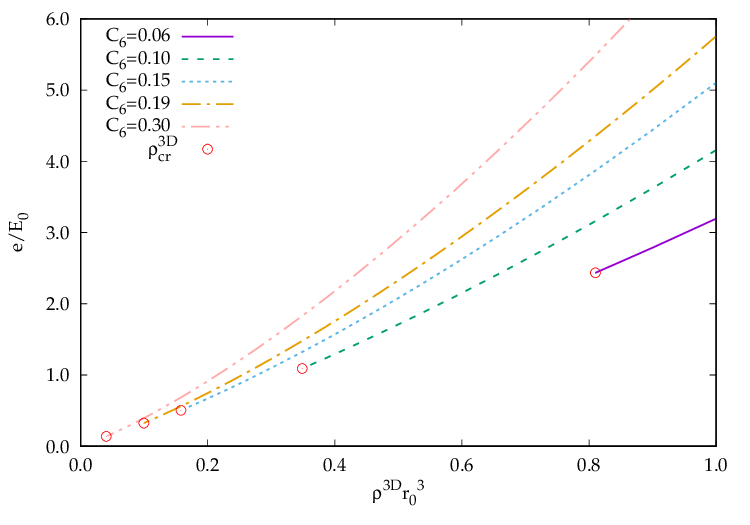}
    \caption{Equation of state $e(\rho^{\rm 3D})$ of the homogeneous molecular BEC for several values of the
    repulsion strength $C_6$. The absence of a local
    minimum shows that the molecular BEC is in a gas-like phase, not in a self-bound liquid phase. For each $C_6$,
    $e(\rho)$ ends at a critical density $\rho_{\rm cr}^{\rm 3D}(C_6)$ (open circle),
    where the homogeneous phase becomes unstable against density wave and enters a layered phase.}
    \label{fig:erho3D}
\end{figure}

In Fig.\ref{fig:erho3D}, we show, for a wide range of $C_6$ values, the equation of
state $e(\rho^{\rm 3D})$ of the homogeneous 3D molecular BEC, which bears the hallmarks of a
gas phase: $e(\rho^{\rm 3D})$ is positive and monotonously falling with $\rho^{\rm 3D}$.
However, the equation of state $e(\rho^{\rm 3D})$ does not extend all the way
to $\rho^{\rm 3D}=0$, but ends at a finite critical densities $\rho_{\rm cr}^{\rm 3D}$, indicated by
circles. In other words, the if the molecular BEC in this gas-like phase is allowed to expand,
it cannot expand to arbitrarily small density like a normal Bose-condensed quantum gas. Instead,
at the critical density $\rho_{\rm cr}^{\rm 3D}$, the homogeneous phase of the molecular
BEC becomes unstable and undergoes a phase transition. Numerically, it is manifested
by failing to converge the HNC0-EL equations (\ref{eq:EL})
to a stable solution $g(\qr)$ with the correct behavior $g(\qr)\to 1$ for $r\to \infty$.

\begin{figure}[hbtp]
    \centering
    \includegraphics[width=1.0\linewidth]{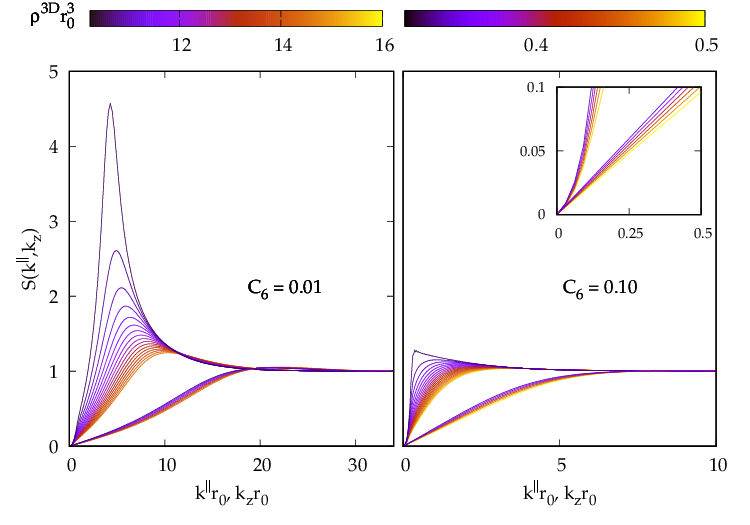}
    \caption{The static structure factor $S(k^\|,k_z)$ in the $z$-direction and parallel direction
    for $C_6=0.01$ (left panel) and $C_6=0.10$ (right panel). The steeper curves correspond to $S(k^\|=0,k_z)$.
    The density is decreasing towards the respective critical density
    $\rho_{\rm cr}^{\rm 3D}$.
    The phase transition from 3D dipole gas to quasi-2D layered fluid is signified by the growing
    peak around in $z$-direction. The inset zooms into the small $\qk$ region for $C_6=0.10$.}
    \label{fig:Sk3D}
\end{figure}

The nature of this phase transition becomes clear if we look at the static structure factor $SS(k^\|,k_z)$.
In Fig.\ref{fig:Sk3D} we show
$S(k^\|,k_z)$ along the $z$-direction and the parallel direction for $C_6=0.01$ (left panel)
and $C_6=0.10$ (right panel). The density (indicated by color) is reduced towards the respective critical density
$\rho_{\rm cr}^{\rm 3D}$. In all cases $S(k^\|=0,k_z)$ are the curves which appear steeper for smaller $k_z$ and
lie above $S(k^\|,k_z=0)$. $S(k^\|=0,k_z)$ develops
a pronounced peak as we approach $\rho_{\rm cr}^{\rm 3D}$ from above, while $S(k^\|,k_z=0)$ changes rather little
over the respective range of densities. If we try to reduce $\rho^{\rm 3D}$ further,
no stable solution to the HNC0-EL equation (\ref{eq:EL}) can be found as mentioned above.
While it cannot be ruled out completely that
this is just a problem of numerical convergence, the fast growth of the peak in $S(k^\|=0,k_z)$ is a rather
strong evidence that density fluctuations in $z$-direction become soft and drive an stability to
a density-modulated phase with a wavenumber $k_{\rm cr}$ given by the peak position. This indicates that
the homogeneous molecular BEC splits
into layers. This effect has been found previously for other dipolar Bose gases~\cite{maciaPRL12,staudingerPRA23}.
We study these layers in section~\ref{ssec:result2D}.

Fig.\ref{fig:Sk3D} might give the impression that in the limit of $\qk\to 0$, $S(k^\|=0,k_z)$ and $S(k^\|,k_z=0)$
have different slopes.
To show that this is not the case, the inset zooms into $S(k^\|,k_z)$ for small wave numbers for $C_6=0.10$.
Both $S(k^\|=0,k_z)$ and $S(k^\|,k_z=0)$ follow eq.(\ref{eq:Skzero}) with the {\em same} speed of sound $c$
in all directions.
The speed of sound, which does not diverge or vanish as we approach $\rho_{\rm cr}^{\rm 3D}$, must be isotropic
in a homogeneous phase, despite the anisotropic interaction (\ref{eq:vbare}).

%rho3D=10.551715425:
\begin{figure}[hbtp]
    \centering
    \includegraphics[width=1.0\linewidth]{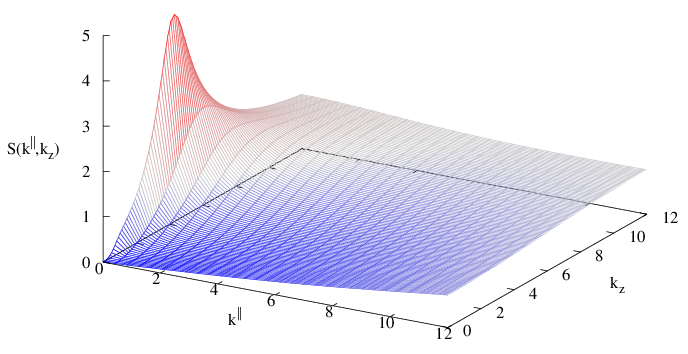}
    \caption{The full static structure factor $S(k^\|,k_z)$ for $C_6=0.01$ and $\rho_{\rm 3D}r_0^3=10.55$,
    corresponding to $S(0,k_z)$ with the highest peak in Fig.\ref{fig:Sk3D} and
    very close to the critical density. The peak is very narrow in $k^\|$-direction.}
    \label{fig:Skmap}
\end{figure}

The peak in $S(k^\|,k_z)$ near $\rho_{\rm cr}^{\rm 3D}$ is sharply localized in $z$-direction.
This can be seen in Fig.~\ref{fig:Skmap} where we show the full anisotropic $S(k^\|,k_z)$ for
$C=0.01$ and $\rho_{\rm 3D}r_0^3=10.55$, corresponding to $S(0,k_z)$ with the highest peak in the
left panel of Fig.\ref{fig:Sk3D}. Just slightly away from the $z$-axis, the peak vanishes.
Resolving this sharp feature in $k^\|$-direction numerically requires exceedingly large
computational domains in $r^\|$-direction. This finite size effect may lead to some uncertainty in
determining the precise value of $\rho_{\rm cr}^{\rm 3D}$.

%\begin{figure}[hbtp]
%    \centering
%    \includegraphics[width=1.0\linewidth]{apot_over_rho_3D.eps}
%    \caption{Phase diagram showing the phase transition line $(\rho,C_6)$ which separates the
%    homogeneous gas phase from the layered phase.}
%    \label{fig:phasediagram3D}
%\end{figure}
%\begin{figure}[hbtp]
%    \centering
%    \includegraphics[width=1.0\linewidth]{apot_over_rho_3D_neu.eps}
%    \caption{Phase diagram showing the phase transition line $(\rho,C_6)$ which separates the
%    homogeneous gas phase from the layered phase.}
%   \label{fig:phasediagram3D}
%\end{figure}
\begin{figure}[hbtp]
    \centering
    \includegraphics[width=1.0\linewidth]{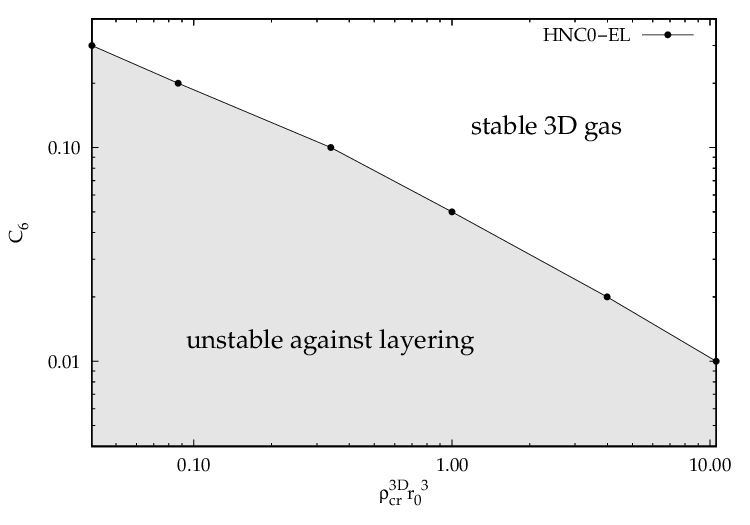}
    \caption{Phase diagram showing the phase transition line $(\rho,C_6)$ which separates the
    homogeneous gas-like phase from the layered phase.}
    \label{fig:phasediagram3D}
\end{figure}

We determined the dependence of the critical density $\rho_{\rm cr}^{\rm 3D}$ on the repulsion parameter $C_6$
in the range $C_6=0.01\dots 0.30$.
%As mentioned above, finding the precise value of $\rho_{\rm cr}^{\rm 3D}$
%is complicated due to numerical finite size effects, but owing to the approximations inherent in the HNC0-EL
%method, the precise value would be affected by these approximations, anyway.
The numerical violation of the sequential relation
\begin{align}
    1 - \rho\int\! d^3r\, (g(\qr)-1) = 0
\end{align}
provides a well-defined instability criterion. Violations larger than $\epsilon=10^{-3}$ are typically
accompanied by non-decaying oscillations of $g(r^\|=0,z)$ for large $z$. Therefore we use this value of $\epsilon$ to
define $\rho_{\rm cr}^{\rm 3D}$.

In Fig.\ref{fig:phasediagram3D} we show the full phase diagram $(\rho^{\rm 3D},C_6)$. The filled circles mark the
critical density $\rho_{\rm cr}^{\rm 3D}$ for a given value of $C_6$, connected by lines to guide the eye.
%The green line is a fit $C_6={a\over (\rho_{\rm cr}^{\rm 3D})^\alpha}$ with exponent $\alpha=0.61\approx {3\over 5}$.
%We see that the HNC0-EL results can be approximated by this power law quite well.
The lines separates the stable 3D gas-like phase from the layer phase, the latter is indicated by the gray area.
For higher values of the repulsion strength $C_6$, that is, for weaker attraction (see Fig.~\ref{fig:pot3D}),
the homogeneous 3D phase becomes more stable and the critical density $\rho_{\rm cr}^{\rm 3D}$ decreases.
Note that there seems to be no maximal value of $C_6$ beyond which the 3D gas phase is unconditionally
stable for all densities down to zero.

%{\color{red}power law?!}

%We estimate the uncertainty
%by calculating $\rho_{\rm cr}^{\rm 3D}$ for several choices of mesh size for $C=0......$ and show the
%variance of the resulting $\rho_{\rm cr}^{\rm 3D}$ as error bar.  Considering the methodological errors due
%to the approximation of HNC0-EL method (only two-body correlations $u(\qr)$ and omission of elementary diagrams),
%the finite size error is not significant.

\subsection{Quasi-two-dimensional liquid layers}
\label{ssec:result2D}

The results of the previous section suggest that the molecular BEC forms layers rather
than a homogeneous gas phase if there is no confinement and the density is allowed to fall.
Here we study a single layer with the method described in section~\ref{ssec:theory2D}. In the following discussion,
we take the direction perpendicular to the layer plane as the $z$-direction.

A layer is stable if it does not disperse in $z$-direction to a homogeneous gas, hence it is
self-bound with respect to the $z$-direction. It can still be a 2D gas with respect to the $xy$-plane, i.e.\
it may lower its energy by dispersing in the $xy$-plane like a gas if there is no confining potential. Alternatively,
it can be a {\em fully} self-bound liquid such that it does not disperse in any direction. In that case a finite
molecular BEC will form a liquid droplet, with thickness $\sigma$ and an approximately
constant area density inside the droplet, given by the equilibrium density.
We know already that such droplets are possible from
exact DMC simulations~\cite{langenPRL25}.

\begin{figure}[hbtp]
    \centering
    \includegraphics[width=1.0\linewidth]{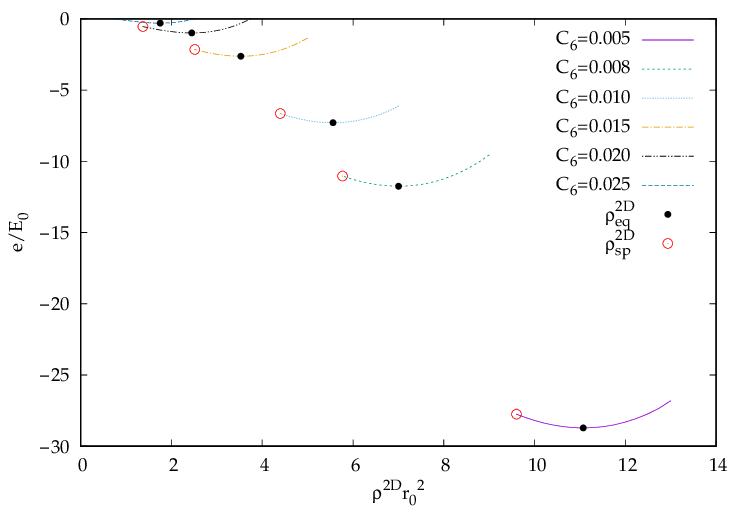}
    \caption{Equation of state $e(\rho^{\rm 2D})$ of a layer of the homogeneous molecular BEC for several
    values of the repulsion strength $C_6$. The local minimum (filled circles) shows that the layer is not
    only self-bound with respect to the $z$-direction, but is indeed a fully self-bound liquid for these
    and smaller values of $C_6$,
    with equilibrium density $\rho^{\rm 2D}_{\rm eq}$ attained at the minimum of $e(\rho^{\rm 2D})$.
    $e(\rho^{\rm 2D})$ ends at the spinodal density $\rho_{\rm sp}^{\rm 2D}$ (open circles),
    where a homogeneous layer becomes unstable against phase separation.}
    \label{fig:erho2D}
\end{figure}

We study the layer phase systematically for a wide range of repulsion
parameters $C_6$ by calculating the 2D equation of state $e(\rho^{\rm 2D})$ where $\rho^{\rm 2D}$ is
the area density, $\rho^{\rm 2D}=\int dz\, \rho^{\rm 3D}(z)$. $\rho^{\rm 3D}(z)$ has a Gaussian
profile owing to the ansatz (\ref{eq:Psi2D}) for the many-body ground state, where the width $\sigma$
is determined by solving (\ref{eq:dEdsigma}) self-consistently.

Fig.~\ref{fig:erho2D} shows $e(\rho^{\rm 2D})$ for several repulsion strengths
between $C_6=0.005$ and $C_6=0.025$. We observe that $e(\rho^{\rm 2D})$ has a minimum, hence
it is a liquid for these (and smaller) values of $C_6$. The respective equilibrium densities
$\rho^{\rm 2D}_{\rm eq}$ are shown as filled circle.
Without any external potential such as a trapping potential, the system
will adjust itself to this density; in case of a large, but finite system, it will form a droplet with a central
density of $\rho^{\rm 2D}_{\rm eq}$, as mentioned above. Hence $\rho^{\rm 2D}_{\rm eq}$ is the
relevant density for experiments, where a finite cloud of molecules is prepared.

For $C_6 > 0.035$, we still find stable
layer solutions, but $e(\rho^{\rm 2D})$ stays positive for all $\rho^{\rm 2D}$ and it
has no minimum. In other words, when the dipole-interaction is strongly shielded,
the attraction becomes too weak to support a quasi-2D liquid. The molecular BEC
is a quasi-2D gas, although it is still self-bound with respect to the $z$-direction.
%Such a gaseous layer can only be metastable because the system can lower its energy towards zero
%by expanding like a gas also in $z$-direction.
Between $C_6=0.028$ and $C_6=0.035$, we have the peculiar situation that
$e(\rho^{\rm 2D})$ is positive, but still has a local minimum at a finite equilibrium density.
This constitutes a meta-stable liquid layer.

Even for $C_6 < 0.28$, $e(\rho^{\rm 2D})$ eventually becomes positive if we keep increasing $\rho^{\rm 2D}$,
until we reach an upper critical density where stable solutions with a finite width $\sigma$ cease
to exist. $e(\rho^{\rm 2D})>0$ again implies a metastable
layer because the system can lower its energy, e.g.\ by rearranging into two well-separated layers with
area density ${\rho^{\rm 2D}\over 2}$.

Without a general pair correlation $u(z_1,z_2,r_{12}^\|)$, our ansatz (\ref{eq:Psi2D})
may be too simply to answer whether metastable liquid or gaseous layers exist.
In Fig.~\ref{fig:erho2D} we therefore show only results with negative energy, and only in the
vicinity of the equilibrium density.  But the lower end of the $e(\rho^{\rm 2D})$ curve,
marked by an open circle, does have a physical meaning: this is a prediction of
the spinodal density $\rho^{\rm 2D}_{\rm sp}$.
As for all liquids, the equation of state ends at $\rho^{\rm 2D}_{\rm sp}$.
At $\rho^{\rm 2D}_{\rm sp}$ the homogeneous liquid layer becomes
unstable against long-wavelength density fluctuation and undergoes a phase transition
to a two-phase mixture of liquid and gas. At $\rho_{\rm sp}^{\rm 2D}$, the compressibility
$\kappa={d\rho\over d P}$ diverges. The inverse of the compressibility is related to the speed of sound $c$
\begin{align}
    {1\over\kappa}=mc^2={d\over d\rho}\rho^2{de\over d\rho}
    \label{eq:mc2}
\end{align}
We can determine ${1\over\kappa}$ either via $c$ from the slope of $S(k)$ according to eq.(\ref{eq:Skzero}),
or from its thermodynamic definition according to eq.(\ref{eq:mc2}). In an exact implementation
of the HNC-EL method, the two ways to obtain $mc^2$ would give the same answer.
Due to the approximations (omitting the elementary diagrams and pair correlations that decouple
parallel and perpendicular degrees of freedom) the two methods will given different values for $mc^2$, and
therefore different $\rho_{\rm sp}^{\rm 2D}$. In the appendix we compare $mc^2$ results
obtained with the two methods, which shows larger deviations for larger density, as expected.
For Fig.~\ref{fig:erho2D}, we use the slope of $S(k)$ to determine
$\rho_{\rm sp}^{\rm 2D}$, because this is where the numerical procedure to solve
eqns.(\ref{eq:dEdsigma}) and (\ref{eq:dEdu2D}) stops to converge -- the
procedure does not give numerical solutions when there is no physical homogeneous solution.
We note that even with fewer approximations in the HNC-EL method, determining the precise spinodal density is
challenging~\cite{Campbell96}, and the same is true for quantum Monte Carlo methods due to large finite size
effects near second order phase transitions. Since experimental determination of $\rho_{\rm sp}^{\rm 2D}$
would be very hard, its accurate theoretical determination is not our priority.

\begin{figure}[hbtp]
    \centering
    \includegraphics[width=1.0\linewidth]{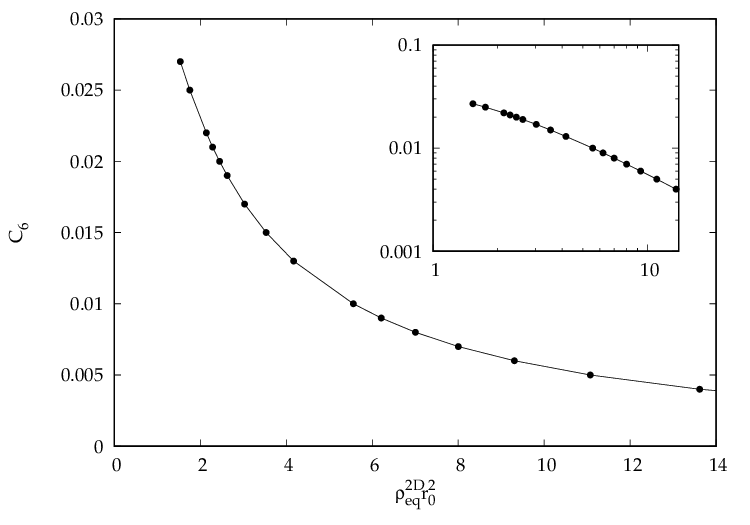}
    \caption{Relation between equilibrium density $\rho_{\rm eq}^{\rm 2D}$ and repulsion parameter $C_6$.
    The same plot on logarithmic scales (inset) shows there is no simple power law describing the relation.}
    \label{fig:rhoeq_neu}
\end{figure}

\begin{figure}[hbtp]
    \centering
    \includegraphics[width=1.1\linewidth]{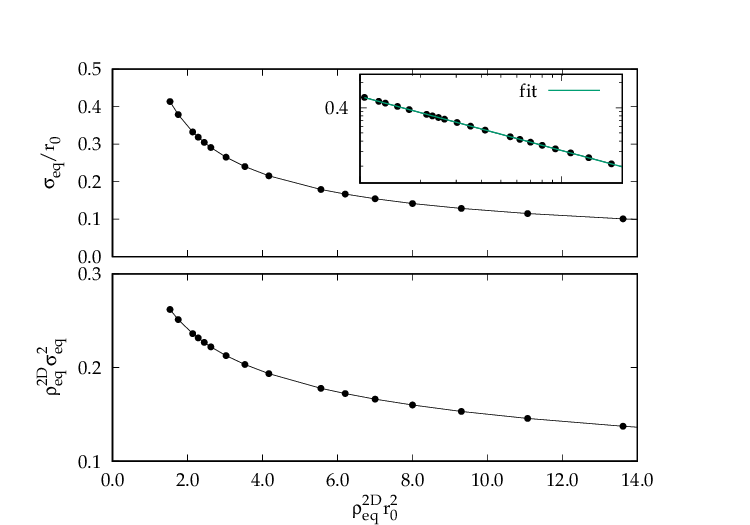}
    \caption{Top panel:
    equilibrium width $\sigma_{\rm eq}$ of a self-bound
    layer as function of $\rho_{\rm eq}^{\rm 2D}$. The double logarithmic
    representation in the inset shows that $\sigma_{\rm eq}$ follows a power
    law with an approximate exponent ${2\over 3}$.
    Bottom panel:
    $\rho_{\rm eq}^{\rm 2D}\sigma_{\rm eq}^2$ as function of $\rho_{\rm eq}^{\rm 2D}$.}
    \label{fig:sigma}
\end{figure}

The width $\sigma$ of a layer is obtained by
minimizing the energy. In the top panel of Fig.\ref{fig:sigma}, we show the width $\sigma_{\rm eq}$
at the equilibrium density $\rho_{\rm eq}^{\rm 2D}$. Note that $\sigma_{\rm eq}$
{\em falls} with increasing $\rho_{\rm eq}^{\rm 2D}$ (i.e. decreasing repulsion $C_6$, leading
to a more attractive interaction).
$\sigma_{\rm eq}(\rho_{\rm eq}^{\rm 2D})$ follows very closely a power law, as can be
seen in the inset of Fig.\ref{fig:sigma},
\begin{align}
    \sigma_{\rm eq} \approx {b\over (\rho_{\rm eq}^{\rm 2D})^\beta}
\end{align}
with an exponential $\beta=0.65\approx {2\over 3}$. The quality of the fit is remarkable, however,
the density profile is based on the simple Gaussian decoupling ansatz (\ref{eq:Psi2D}).
Furthermore, the approximations inherent in the HNC0-EL method lead to larger errors for larger
$\rho_{\rm eq}^{\rm 2D}$. The exact layer width may deviate from our result and may not
follow a power law.
%The linear fit is $y=-0.649x-0.607$

As mentioned above, the ansatz (\ref{eq:Psi2D}) is only justified if $\sigma$
is sufficiently small compared to the average interparticle distance.
In the bottom panel of Fig.\ref{fig:sigma}, we see that $\rho_{\rm eq}^{\rm 2D}\sigma_{\rm eq}^2<1$
for all equilibrium densities, but not by orders of magnitude.

\section{Conclusions}
\label{sec:conclusion}

%We appear to have the interesting case of a 2D fluid which has all the hallmarks of a liquid:
%it can minimize its energy by equilibrating to the equilibrium density $\rho_{\rm eq}^{\rm 2D}$, and if the density is
%reduced, it phase separates at the spinodal density $\rho_{\rm sp}^{\rm 2D}$. However,
%if the density is increased, it does not become a solid, but is loses cohesion at a critical
%density $\rho_{\rm cr}^{\rm 2D}$ and becomes a 3D gas. {\color{red} BUT this happens at POSITIVE energy which
%makes no sense!}

We studied a microwave-shielded molecular Bose Einstein condensate interacting via effective dipole-dipole
interactions.%, which is the reverse of the usual dipole-dipole interaction: repulsive for head-to-tail
%configuration (taken to be the $z$-direction), and attractive in a plane perperpendular to the $z$-direction.
For our calculation we use a variational many-body method, the hypernetted-chain Euler-Lagrange method,
which accounts for correlations non-perturbatively \cite{Kro86,KroTrieste,QMBT00Polls}.

For a given density $\rho^{\rm 3D}$, a homogeneous phase of such an anisotropic BEC is stable for 
strong enough shielding of the dipolar attraction, corresponding to a sufficiently large
repulsion parameter $C_6$ in the effective interactions (\ref{eq:vbare}).
Or conversely, for a given $C_6$, it is stable for a large enough $\rho^{\rm 3D}$.
However, if we decrease $\rho^{\rm 3D}$, the homogeneous phase becomes unstable against density
waves in $z$-direction at a critical density $\rho^{\rm 3D}_{\rm cr}$.
This can be seen from the growing peak in the anisotropic static structure factor
$S(\qk)$ for $\qk$ pointing in the $z$-direction, the direction of dipolar repulsion.
We calculated the full phase diagram (Fig.\ref{fig:phasediagram3D})
to determine the critical density $\rho^{\rm 3D}_{\rm cr}$ as function of $C_6$.
We find no upper bound of $C_6$ beyond which it would be unconditionally stable.

Instabilities against a density wave has been predicted for other dipolar
systems~\cite{maciaPRL12,staudingerPRA23}. The present case is peculiar, however:
on the one hand, the equation of state $e(\rho^{\rm 3D})$ appears gas-like, because
$e(\rho^{\rm 3D})$ is positive and monotonously falling with $\rho^{\rm 3D}$.
On the other hand $e(\rho^{\rm 3D})$ stops at the critical density $\rho^{\rm 3D}_{\rm cr}$:
we can adiabatically expand the molecular BEC like a gas until it reaches
$\rho^{\rm 3D}_{\rm cr}$ where it enters the layer phase.
Therefore, the system is not a true quantum gas.

Hence, without a confining potential,
the layer phase is the equilibrium state of a microwave-shielded molecular BEC.
We studied a single layer with the HNC0-EL method with the simplification of a
Gaussian density profile and decoupling parallel and perpendicular degrees of freedom.
The layer width $\sigma$ is determined self-consistently according to Ritz'
variational principle. We find that such quasi-2D
layers are not only stable for a wide range of area densities $\rho^{\rm 2D}$,
but are {\em self-bound} for $C_6\lesssim 0.028$, i.e.\ if the dipole-dipole attraction is not
not too strongly shielded. These self-bound layers have all the characteristic properties of a liquid:
({\it i\/}) the energy per particle $e(\rho^{\rm 2D})$ is negative;
({\it ii\/}) $e(\rho^{\rm 2D})$ has a minimum, defining the equilibrium density $\rho^{\rm 2D}_{\rm eq}$;
({\it iii\/}) $e(\rho^{\rm 2D})$ ends at the spinodal density $\rho^{\rm 2D}_{\rm sp}$, where
the compressibility diverges and the layer phase separates into liquid and gas.
The strength of the repulsion $C_6$ can be varied
over a wide range by choosing the Rabi frequency and the detuning appropriately\cite{langenPRL25}.
Fig.\ref{fig:rhoeq_neu} shows that this in turn means that $\rho^{\rm 2D}_{\rm eq}$ varies
over a wide range.

In all studied cases, we find that the layer width at equilibrium $\sigma_{\rm eq}$ is
smaller than the average inter-particle distance, $\rho_{\rm eq}^{\rm 2D}\sigma_{\rm eq}^2<1$.
These quasi-2D layers can be regarded as essentially two-dimensional, and this also
validates the decoupling approximation.
Our results thus show that molecular Bose gases offer the opportunity to study
two-dimensional quantum liquids over a wide range of densities, probably spanning
several orders of magnitude.

\section{Outlook}
\label{sec:outlook}

The field of quantum gases of microwave-shielded molecules is very young, and the family of HNC-EL methods can
shed light on many interesting questions about stability, structure and dynamics of molecule quantum gases.

In this work we studied only a single, isolated layer, but the transition from the 3D gas phase to
the liquid layer phase may result in multiply layers. The long range of the dipole-dipole interaction
means that these layers are dynamically coupled. The resulting inter-layer correlations and their effects
have been studied in the past\cite{raderPRA17,hebenstreitPRA16}. In particular, HNC-EL calculations
of bilayers of anti-parallel dipoles have predicted self-bound correlated bilayer liquid phases
\cite{hebenstreitPRA16}. These predictions will change for microwave-shielded molecules,
where the dipole interaction is essentially inverted, and already a single layer is self-bound. 

As mentioned in the introduction, dipolar quantum gases can support roton excitations,
which are precursors to a softening of an elementary excitation with finite momentum
\cite{Santos2003,odellPRL03,chomazNatPhys18,Natale2019,Schmidt2021,Blakie2020}.
The time-dependent extension of the HNC-EL method in combination with linear response theory,
the correlated basis function (CBF) method\cite{Saarela86,Clements93,apajaPRL03,krotscheckPRB08,zillichJCP10b,campbellPRB15},
has been used to study the dynamics of dipolar quantum gases
\cite{dipolePRL09,hufnaglJLTP10,hufnaglPRA13,raderPRA17}, including a roton-roton crossover
in trapped layers\cite{hufnaglPRL11}. The high peak in $S(\qk)$ of the 3D gas phase in this work
suggests that a roton mode appears close to the phase transition to the layer phase. We will adapt
the CBF method to calculate the dynamic structure function $S(\qk,\omega)$ in the 3D phase
and in the layer phase, and obtain dispersions relations of elementary excitations and their damping.
With the recently developed time-dependent HNC-EL method\cite{gartnerSciPost25}, nonequilibrium
many-body dynamics beyond the linear response regime can be studied, e.g.\ actuated by modulated
microwave fields, leading to a time-dependent effective dipole interactions (\ref{eq:vbare}).

The effective interaction (\ref{eq:vbare}) with cylinder symmetry represents the
special case of shielding with circularly polarized microwaves. In the general case
of elliptic polarization, the resulting effective interaction is
anisotropic in all three directions\cite{jinPRL25}. The many-body physics
of such an interaction has not been explored to our knowledge. Preliminary HNC-EL results
indicate that self-bound layers are quite robust and persist also for fully anisotropic
interactions\cite{polterauerinprep}.

Our Jastrow ansatz for the wave function for the quasi-2D layer (\ref{eq:Psi2D}) decouples the in-plane and out-of-plane
motion, and accounts only for in-plane correlations. This is a good approximation for strong self-confinement,
where the layer thickness $\sigma$ is much smaller than the average particle distance $(\rho^{\rm 2D})^{-1/2}$.
We do find that $\rho^{\rm 2D}\sigma^2<1$ near the equilibrium density, but not by an order of magnitude,
see Fig.\ref{fig:sigma}.
Hence the decoupling approximation may lead to a bias in our results.
The Jastrow ansatz can be improved to general correlations $u(|\qr_1^\|-\qr_2^\|,z_1,z_2$, as mentioned in
section~\ref{ssec:theory2D}, and the Gaussian approximation for the density profile can be generalized
to a functional optimization of $\phi(z)$. Instead of the EL-equations (\ref{eq:dEdsigma}) and (\ref{eq:dEdu2D}),
the following Euler-Lagrange equations have to be solved self-consistently:
\begin{align}
    {\delta e\over \delta\phi(z)} &= 0 \\
    {\delta e\over \delta u(r^\|,z_1,z_2)} &= 0
\end{align}
These inhomogeneous HNC-EL equations are much more complicated, but have been applied successfully
to predict the structure and layering transitions of thin superfluid $^4$He films adsorbed on
substrates\cite{EpKro,Clements93,clementsPRB96} -- indeed, there are similarities between the physics
of $^4$He films and of self-bound dipolar layers. We expect that the inhomogeneous HNC-EL method
is better suited to describe the details of the transition between liquid layers and the anisotropic homogeneous
3D phase, and we will pursue this direction in our further studies.

\begin{acknowledgments}
We thank Ferran Mazzanti for fruitful discussions.
\end{acknowledgments}

\begin{appendix}

\section{Compressibility and spinodal point}

At the spinodal point the compressibility diverges. We can calculate the spinodal
density $\rho^{\rm 2D}_{\rm sp}$ by finding the zero of the inverse
compressibility $mc^2$ as function of $\rho^{\rm 2D}$. Compressibility is a measure
of the susceptibility of the layer to long-wave density fluctuations in parallel
direction. $mc^2$
can be calculated in (at least) two ways, and only an exact theory guarantees that both
ways yield the same result. Approximate theories like the HNC0-EL method generally
give inconsistent results. The discrepancy between the two method can be regarded
as a quality check (and a guide to improve the method by reducing the discrepancy).

\begin{figure}[hbtp]
    \centering
    \includegraphics[width=1.0\linewidth]{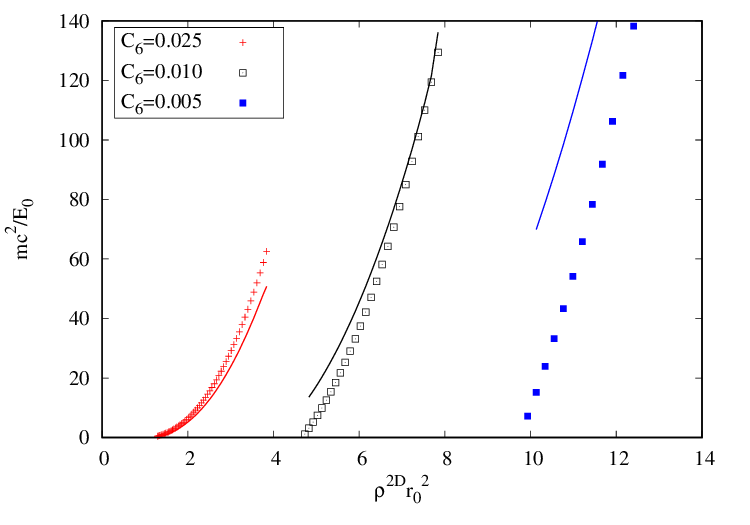}
    \caption{The inverse compressibility $mc^2$ of a layer as function of density $\rho^{\rm 2D}$ for
    $C_6=0.025; 0.010; 0.005$, calculated from the slope of $S(k)$ according to
    eq.~(\ref{eq:Skzero}) (symbols) and from the thermodynamic expression eq.~(\ref{eq:mc2})
    (lines with the same color). For small $C_6$, i.e. strong attraction, the thermodynamic
    $mc^2$ curve is shifted significantly to lower density, and would predict a lower spinodal
    density $\rho^{\rm 2D}_{\rm sp}$, where $mc^2\to 0$.}
    \label{fig:mc2compare}
\end{figure}

In Fig.\ref{fig:mc2compare}, we compare $mc^2$ as function of density $\rho^{\rm 2D}$
for three values of the repulsion parameter $C_6$. The symbols are the results obtained
from the slope of $S(k)$ (\ref{eq:Skzero}),
and the lines are the results obtained from the thermodynamic expression (\ref{eq:mc2}).
Equal colors indicate equal $C_6$ values.
With decreasing $C_6$, the attraction becomes stronger
and the relevant density range increases, see also Fig.\ref{fig:erho2D}, and the discrepancy
between the two $mc^2$ estimates increases as well. This trend can be understood from
the increased importance of the elementary diagrams with increasing density\cite{Kro86}.
We conclude that in the regime of high-density liquids for
$C_6=0.005$ and lower, elementary diagrams become important for quantitative results;
we remind that at high density, triplet correlations $u_3(\qr_1,\qr_2,\qr_3)$ become important,
too.

\end{appendix}

\bibliography{zotero,my,bec,ocshehy,papers,dipoles}

\end{document}